# Defect Engineering for Modulating the Trap States in Two-dimensional Photoconductor

*Jie Jiang, Chongyi Ling, Tao Xu, Wenhui Wang, Xianghong Niu, Amina Zafar, Zhenzhong Yan, Xiaomu Wang, Yumeng You, Litao Sun, Junpeng Lu, Jinlan Wang[*] and Zhenhua Ni[*]*

J. Jiang, C. Ling, Dr. W. Wang, Dr. A. Zafar, Z. Yan, Prof. J. Lu, Prof. J. Wang, Prof. Z. Ni

School of Physics and Key Laboratory of MEMS of the Ministry of Education, Southeast University

Nanjing, 211189, China.

E-mail: Z. N. (zhni@seu.edu.cn) and J. W. (jlwang@seu.edu.cn).

Dr. T. Xu, Prof. L. Sun

SEU-FEI Nano-Pico Center, Key Laboratory of MEMS of Ministry of Education, Southeast University

Nanjing, 210096, China.

Dr. X. Niu

School of Science, Nanjing University of Posts and Telecommunications

Nanjing, 210046, China.

Prof. X. Wang

School of Electronic Science and Technology, Nanjing University

Nanjing, 210093, China

Prof. Y. You

Ordered Matter Science Research Center, Southeast University

Nanjing, 211189, China.

**Abstract**

Defect induced trap states are essential in determining the performance of semiconductor photodetectors. The de-trap time of carriers from a deep trap could be prolonged by several orders of magnitude as compared to shallow trap, resulting in additional decay/response time of the device. Here, we demonstrate that the trap states in two-dimensional $ReS_2$ could be efficiently modulated by defect engineering through molecule decoration. The deep traps that greatly prolong the response time could be mostly filled by Protoporphyrin ($H_2PP$) molecules. At the same time, carrier recombination and shallow traps would in-turn play dominant roles in determining the decay time of the device, which can be several orders of magnitude faster than the as-prepared device. Moreover, the specific detectivity of the device is enhanced (as high as ~$1.89 \times 10^{13}$ Jones) due to the significant reduction of dark current through charge transfer between $ReS_2$ and molecules. Defect engineering of trap states therefore provides a solution to achieve photodetectors with both high responsivity and fast response.

The trap states in semiconductor photoconductor play important roles in prolong the lifetime of the photo excited carriers, and hence introduce high photoconductive gain to the system.[1-5] On the other hand, the properties of trap states would strongly influence the response or decay time of the photoconductor. For example, extra decay time comes from the thermal re-excitation of trapped carriers into the conduction or valence band,[1, 2, 6] which could be in the timescale of second or minute for some deep traps.[7-9] To optimize the performance of a photoconductor, one has to look deeply into the role of trap states, and finally modulate their properties to balance the sensitivity and response time. The modulation of trap states in semiconductors is mainly focused on surface/interface treatment,[10-12] which is insufficient for conventional semiconductors, e.g. silicon, because their thicknesses in the third dimension. Two-dimensional (2D) materials then provide an excellent platform to study these problems since they are essentially interface-type materials,[13, 14] and the electronic properties could be flexibly modulated through defect and interface engineering.[15-20]

Here, we show an example that by properly engineering the intrinsic defects in 2D $ReS_2$[21] through molecule decoration, the deep traps that greatly prolong the response time could be mostly passivated. Carrier recombination and shallow traps would in-turn play dominant roles in determining the decay time of the device, which can be several orders of magnitude faster than the as-prepared device. Moreover, the specific detectivity of the phototransitor is greatly enhanced (as high as $10^{13}$ Jones) due to reduction of dark current through charge transfer between $ReS_2$ and molecules.

Defect engineering of trap states therefore provides a solution to achieve 2D photodetectors with both high responsivity and fast response.

The schematic of the band structure of a semiconductor containing trap and recombination centers is shown in **Figure 1**a. For simplicity, only one shallow trap, one deep trap and one recombination center are considered, and the semiconductor is $n$ type. The decay time of the photoconductor can be expressed by[1,2,6] $\tau_{decay} = \tau_r + \tau_t(1+\rho)$, where $\tau_r$ is the recombination time of the excess carriers, which is the lifetime of the carriers, $\tau_t$ is the time required for thermal re-excitation of trapped carriers into the conduction or valence band. $\rho$ is the probability that an electron/hole is re-trapped before recombination. The additional decay time is then determined by the time required for emptying of traps (i.e. the de-trap time) $\tau_t$, and inversely proportional to the thermal emission rate of carriers:[22-24]

$$\tau_t^{-1} = s_p N_v V_{th} \exp(\frac{-\Delta E}{kT}) \tag{1}$$

Where $s_p$ is the capture cross section of the trap center, $N_v$ is the effective density of states in the valence bands, $v_{th}$ is the thermal velocity of the carriers ($v_{th} \propto (kT)^{1/2}$) and $\Delta E$ is the energy difference between the trap state and the valence band edge ($\Delta E = E_{tp} - E_v$). Accordingly, the de-trap time could be differed by several orders of magnitude when the energy difference $\Delta E$ of trap states varied by only ~0.1-0.2 eV. Therefore, a deeper trap with larger $\Delta E$ will greatly increase the response time of the device. The decay time of the device is then schematically drawn in Figure 1b, which contains three procedures: the first stage is the recombination of excess carrier, followed by the empty of shallow and deep traps, respectively. Figure S2 gives the experimentally obtained transient response of as-prepared 3-layer $ReS_2$ photoconductor (detailed sample characterization could be found in Figure S1), which contains three different decay procedures in different timescales (sub-millisecond, sub-second, and minute). This suggests that there are different trap centers with different $\Delta E$, and could be due to the presence of structural defects in 2D $ReS_2$, e.g. S vacancies. Zhang[25] *et.al.* demonstrated that chalcogen vacancies will lead to *n*-doping of $ReS_2$. Horzum[26] *et al.* confirmed that S vacancies are the most commonly existed defects in intrinsic $ReS_2$. Liu[8] *et al.* estimated the density of trap states in as-exfoliated $ReS_2$ by studying the temperature dependence of the field-effect mobility, which gave a value of $1.96 \times 10^{13}$ cm$^{-2}$. Our Transmission Electron Microscopy (TEM) results in Figure S3 also confirm the existence of S vacancies in as-prepared $ReS_2$.

Pure $ReS_2$ monolayer is calculated to be a semiconductor with a direct energy gap of ~1.42 eV (**Figure 2**a), which is in consistent with previous work[21]. The formation of S vacancy would introduce several localized defect states within the bandgap of $ReS_2$ (levels 1-3 in the left panel of Figure 2b). Through the analysis of partial charge density distribution (right panel of Figure 2c), the localized states are contributed by the Re atoms around the S vacancy. These localized states may act as recombination or trap states for photo-excited carriers, depending on their positions in band gap and the Fermi level. As an *n*-type semiconductor, the Fermi level of as-prepared $ReS_2$ is close to conductance band. Under illumination, the localized states between the quasi-Fermi levels of electrons and holes will act as recombination centers and states above/below electron/hole quasi-Fermi level will be traps for electrons/holes (Figure S4).[1, 2, 27] Thus, defect level 1 is apparently shallow traps for photo-excited carriers, while level 2 is deep traps contributing to the ultraslow response of as-prepared $ReS_2$ device (Figure S2) and level 3 could be recombination centers. Therefore, $ReS_2$, as a 2D material containing both shallow and deep traps, is a suitable candidate to demonstrate the strategy of defect engineering for modulating the trap states.

It is then highly desirable to minimize the contribution from deep traps that greatly limit the decay time of $ReS_2$ photoconductor. As can be seen in Figure 2b, the pristine localized defect states disappear after Protoporphyrin ($H_2PP$) decoration and several new energy states (level 1-5 in the left panel of Figure 2c) appear due to the adsorption of $H_2PP$ molecule on the S vacancy. As revealed by partial charge density distribution in the right panel of Figure 2c, level 1-4 belong to $H_2PP$ molecule and are not related to $ReS_2$. This suggests that the deep trap centers in $ReS_2$ can be passivated by molecule decoration (we will show that level 5 is indeed a recombination center in the later section). In addition, Bader charge analysis shows that there is 0.65 $e$ charge transfer from $ReS_2$ to $H_2PP$ molecules (Figure 2d), introducing $p$-doping to $ReS_2$, which is in accordance with the transport measurement in Figure S5a. The adsorption of a free $H_2PP$ molecule on S vacancy of $ReS_2$ only needs to overcome a rather small barrier of 0.60 eV (Figure 2e), demonstrating that $H_2PP$ molecules can be easily adsorbed and chemically bonded on S vacancies of $ReS_2$. The stability of $H_2PP$ molecules on ReS$_2$ is also checked by DFT calculation, as shown in Figure S7.

The transient response of as-prepared and H$_2$PP decorated ReS$_2$ is shown in **Figure 3**. The photoresponse of as-prepared ReS$_2$ is dominant by deep traps and in the timescale of minutes, which is in consistent with previous reports.[8, 9] It should be noted that, when shallow and deep traps are both present, the longer-lived, deeper-lying traps are gradually filled first by photo-excited electrons under light illumination.[28-30] In the presence of large number of deep traps, the quasi Fermi level for holes will be pinned near the deep trap states even under high illumination intensity (Figure S13a).[31] The density of trapped holes $p_t$ can be expressed by:[1, 2]

$$p_t = p(P_t/N_v)\exp(\frac{E_{tp}-E_v}{kT}) = P_t\exp(\frac{E_{tp}-E_{Fp}}{kT}) \qquad (2)$$

It is clearly that the traps near the quasi Fermi level for holes have the highest trapping efficiency. This is the reason why deep traps play a dominant role in as-prepared ReS$_2$ photoconductor. On the other hand, the response time of H$_2$PP decorated ReS$_2$ is 3-4 orders fastened and contains two prominent procedures. According to the schematic description in Figure 1, these two procedures are corresponding to carrier recombination and shallow traps, which are in the timescale of sub-millisecond and sub-second. The emission of holes in shallow traps of H$_2$PP decorated ReS$_2$ is in the same time scale of the second stage of as-prepared ReS$_2$ (Figure S2b). It is reasonable to believe that the shallow traps are the same species come from the residual of S vacancies. The absence of new deep traps indicates that the localized state 5 in Figure 2c is indeed a recombination center. Detailed characterizations of H$_2$PP decorated ReS$_2$ device can also be found in Figure S5 and

S6. In order to verify the existence of shallow trapping centers, temperature dependent photoresponse of $H_2PP/ReS_2$ device is investigated,[22-24] and energy difference of traps and valence band edge ($\Delta E$) of ~0.13~0.17 eV is obtained (Figure S8), which agrees reasonably well with theoretical calculations in Figure 2a (defect level 1 is ~0.2 eV above valance band edge).

The two decay processes can also be seen by the logarithmic plot of the normalized photocurrent in **Figure 4**a, which can be fitted with an exponential function with two relaxation times:

$$I_{ph} = A_1 \exp(-t/\tau_1) + A_2 \exp(-t/\tau_2) \tag{3}$$

The fast and slow components of the decay time extracted at different incident laser powers are shown in Figure S5e and S5f, respectively. The contribution to the photocurrent from recombination of excess carriers and shallow traps is also extracted and shown in Figure 4b. As can be seen, the photocurrent contributed by shallow hole traps rises and saturates at power of ~5 μW, while that contributed by recombination of excess carriers continues to increase with light power. Evidently this saturation is associated with the filling of all the shallow trapping centers. This phenomenon is consistent with J. A. Hornbeck and J. R. Haynes's research in silicon.[28, 29] Since for each trapped hole there is one added mobile majority carrier $P_t = \Delta\sigma/e\mu$, the rest shallow trap density is found to be $10^{10}$ cm$^{-2}$, considering the mobility of electron in $H_2PP$ decorated $ReS_2$ is ~13.1 cm$^2$/Vs (Figure S5a). The evolution of photoconductivity decay in different laser power clearly indicates that the contribution from shallow traps becomes more significant with decreasing laser power, and the highest ratio could be more than 60% at 5 pW (Figure 4c).

Figure 4d shows the specific detectivity (D*) of the device as a function of incident laser power ($P_{in}$):[32]

$$D^* = \frac{\sqrt{AB}R}{\sqrt{\langle i_n \rangle^2}} = \frac{\sqrt{AB}R}{\sqrt{\int_0^B S_n(f)df}} \quad (4)$$

Where A is the device area, B is the bandwidth, $\langle i_n \rangle^2 = \int_0^B S_n(f)df$ is the square noise current and R is the photoreponsivity (detailed information about noise measurement and responsivity can be seen in supporting information Figure S9 and S10). The $H_2PP$ decorated $ReS_2$ device shows a remarkable detectivity as high as ~$1.89 \times 10^{13}$ Jones at ~5 pW, which is superior compared to previous reported $ReS_2$ devices[8, 9] and silicon photodetectors.[33] This is mainly contributed to the significant reduction of current noise after $H_2PP$ decoration (Figure S9a), and also the photoconductive gain from shallow traps. Furthermore, $H_2PP$ decoration will introduce new recombination centers into our device and may also increase the carrier lifetime. For example, the addition of recombination centers formed by S vacancies in CdS increase the carrier lifetime and the photoresponsivity[1]. We have fabricated more than 10 devices with the layer number of $ReS_2$ ranges from 2-5 layers, and all the device show similar high detectivity and fast response (see Figure S11, S12 for results from additional devices).

It should be noted that from our DFT calculations, both shallow and deep traps are filled by molecule decoration, while in our experiment, the contribution of shallow traps plays an important role in photoconductivity gain after molecule decoration.

This is because a small amount of S vacancies still remains in $ReS_2$ and their density is about $10^{10}$ cm$^{-2}$ according to our experiment results. In as-prepared $ReS_2$ with a large density of deep traps ($\sim 10^{13}$ cm$^{-2}$),[8] the photoresponse is dominated by deep traps and negligible shallow traps can capture holes, as shown in Figure S13a. After the removal of most S vacancies, the quasi-Fermi level for holes is able to move down and shallow traps are activated (Figure S13b). An evidence of this assumption is that the decorated device still shows increased photoconductivity after stopping light illumination for the time scale of seconds (Figure S13c). It should be noted that the contribution from deep traps is very small and less than 10％ at very weak power (5 pW), and becomes negligible (<1%) with increased light intensity (Figure S13d). The shallow traps then dominate the photoresponse in $H_2PP$ decorated $ReS_2$ device.

In summary, we have demonstrated that the trap states in 2D $ReS_2$ could be modulated through defect engineering. The deep traps that greatly prolong the photoresponse of the photoconductor could be passivated through molecule decoration. As a result, shallow traps and recombination centers dominate the photoresponse, which greatly improve the response speed as well as the sensitivity of the devices. Our study provides a new way to achieve photodetectors with high responsivity and fast response. This shows advantages as compared to other approaches, e.g. hybrid with QDs[7] and carbon nanotubes,[34] introducing midgaps states,[7] which would either limit the spectral response or strongly prolong the response time. This technique could also be applicable to different 2D materials, e.g. black phosphorus with photoresponse in the infrared region, to realize high performance infrared photodetector.

**Experimental Section**

*Device fabrication and characterization:* Thin $ReS_2$ flakes are exfoliated onto a 300nm $SiO_2$/heavily p-doped silicon ($SiO_2$/Si) substrate. Micro-Raman measurements are carried out using a Horiba HR800 confocal Raman system. The thickness is measured by Atomic Force Microscope (AFM, Asylum Research). Source and drain electrodes are patterned by electron beam lithography (FEI, FP2031/12 INSPECT F50). 5 nm Ni/50 nm Au electrodes are deposited by thermal evaporation (TPRE-Z20-IV), and lift-off processes. $H_2PP$ molecules are dissolved in acetone and then deposited on the surface of the $ReS_2$ by drop casting. The concentration of $H_2PP$ solution is about $10^{-4}$ mol $L^{-1}$. Electrical and photoresponse characteristics of the devices are measured using a Keithley 2612 analyzer under ambient conditions and a 532 nm laser is used as light source. The spectral response is obtained using a supercontinuum light source (SuperK Compact nanosecond kilohertz) with a tunable band pass filter. A digital storage oscilloscope (Tektronix TDS 1012, 100 MHz 1 GS $s^{-1}$) is used to measure the transient photoresponse. Temperature-dependent photoresponse is measured in a Montana Instruments cryostation low-temperature system.

*Density functional theory calculations:* Density functional calculations are performed by using the pseudopotential plane-wave method with projected augmented wave potentials[35] and Perdew-Burke-Ernzerhof-type generalized gradient approximation (GGA)[36] for exchange-correlation functional, as implemented in the Vienna *ab* initio simulation package (VASP).[37] The plane-wave energy cutoff is set to be 400 eV. A vacuum space of 12 Å is used to prevent interlayer interactions. Supercells consisting of 4 × 2 × 1 unit cells of $ReS_2$ are used and the Brillouin zones were sampled by a Monkhorst-Pack k-point mesh with a 2 × 2 × 1 kpoint grid. The convergence threshold was $10^{-5}$ eV and 0.03 eV $Å^{-1}$ for energy and force, respectively.

*TEM characterizations:* Thin $ReS_2$ flakes are exfoliated onto a 300 nm $SiO_2$/heavily p-doped silicon ($SiO_2$/Si) substrate and transferred to a holy carbon-coated copper TEM grid by the isopropyl alcohol (IPA) based transfer method.[38] To remove the polymer residue on the sample, the sample is annealed at 130 $^oC$ in a mixture of hydrogen and argon before TEM. TEM imaging is carried out in an image aberration-corrected TEM (FEI Titan 80–300 operating at 80 kV) and a charge-coupled device camera (2 ×2k, Gatan UltraScanTM 1000) is used for image recording with an exposure time of 1 s.

**Acknowledgements**

J.J. and C.L contributed equally to this work. This work is supported by the National Key Research and Development Program of China (No. 2017YFA0205700, 2017YFA0204800), NSFC (61774034, 61422503, 21525311, 21773027, and 11704068), Jiangsu 333 project (BRA2016353), the open research funds of Key Laboratory of MEMS of Ministry of Education (SEU, China), and the Fundamental Research Funds for the Central Universities. The authors would like to thank Prof Chuanhong Jin and Prof Shuai Dong for their help on this work.

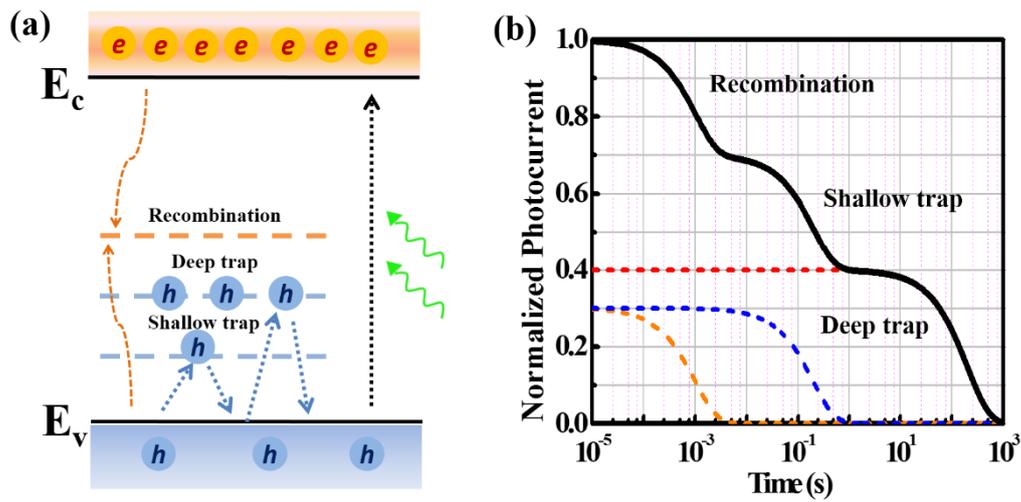

**Figure 1.** Schematic diagram of carrier recombination and trapping kinetics. a) Schematic diagram of the band structure of a semiconductor containing trap and recombination centers. For simplicity, only one shallow trap, one deep trap and one recombination center are considered, and the semiconductor is *n* type. b) Schematically drawn of decay time which contains three procedures.

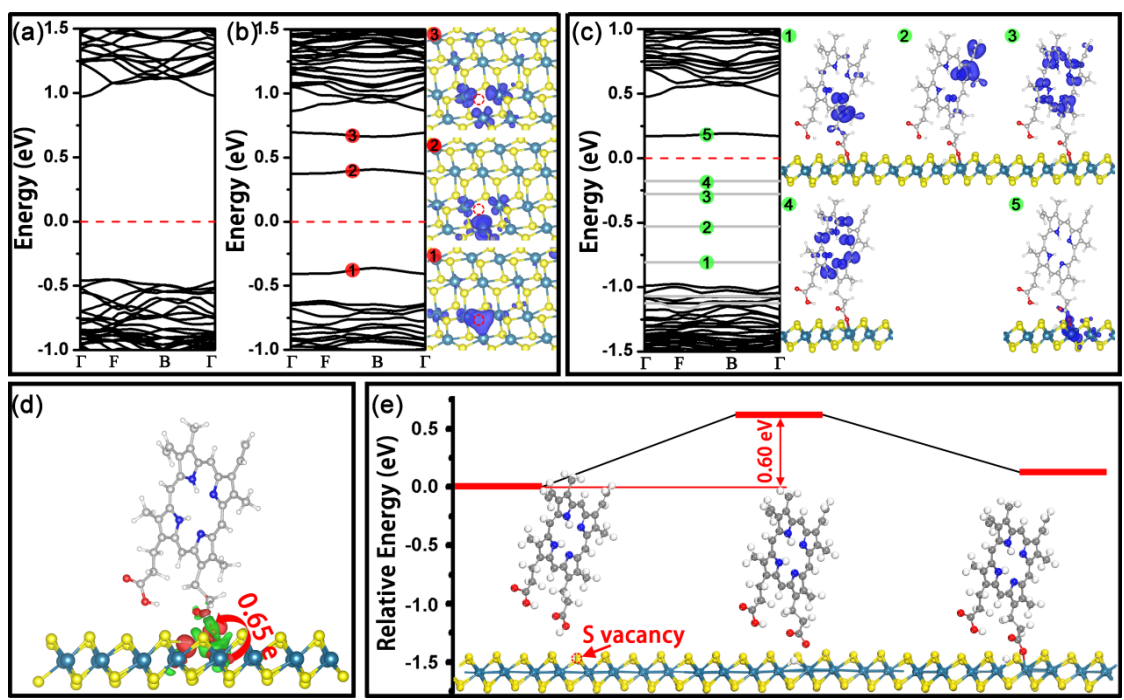

**Figure 2.** Density functional theory calculations. a) Band structures of pristine single layer $ReS_2$. b) Band structures of $ReS_2$ with an S vacancy and the partial charge density distributions of the bands near the Fermi level. c) Band structures of $ReS_2$ with the adsorption of a Protoporphyrin molecule on the S vacancy and the partial charge density distributions of the bands near the Fermi level. d) Charge density difference of a Protoporphyrin molecule adsorbed on S vacancy of $ReS_2$ and the charge transfer is 0.65 $e$ from $ReS_2$ to Protoporphyrin. The positive and negative charges are shown in red and green, respectively. e) Kinetics and transient states of the reaction between a single S vacancy and a Protoporphyrin molecule. The cyan, yellow, gray, white, red and blue balls represent the Re, S, C, H, O and N atoms, respectively.

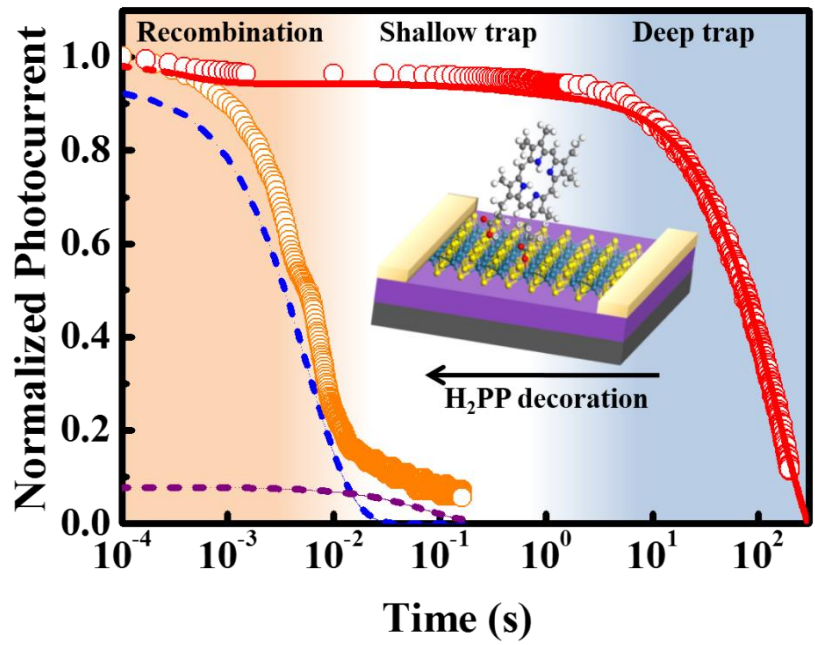

**Figure 3.** Transient response of as-prepared and H$_2$PP decorated ReS$_2$. The deep traps are removed by decoration of H$_2$PP molecules. Inset is the schematic diagram of our device

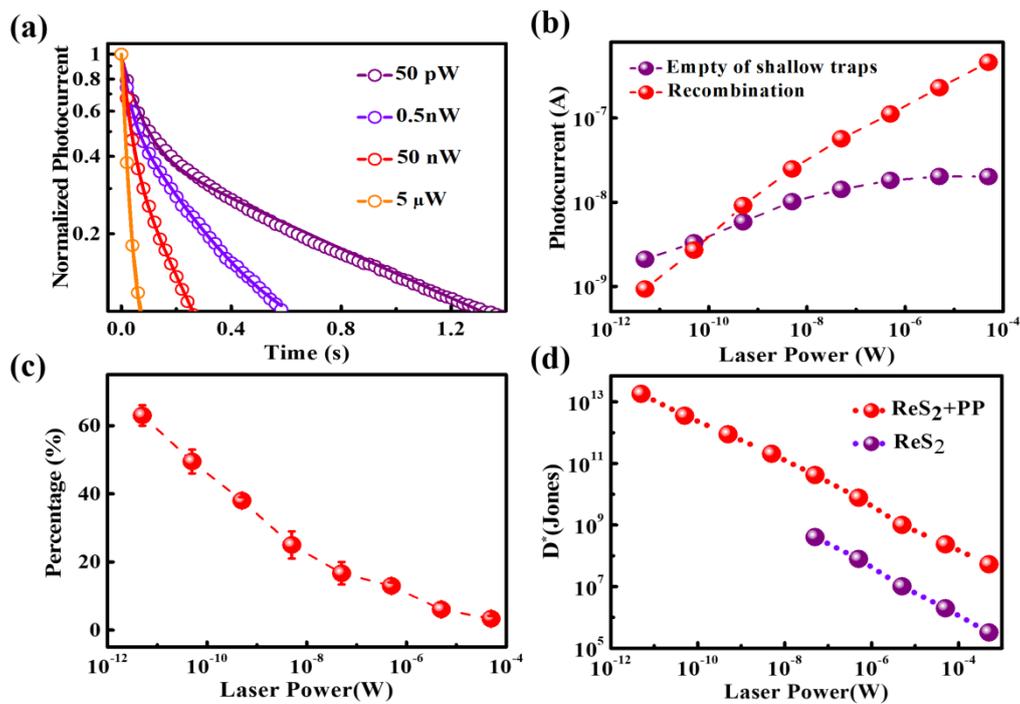

**Figure 4.** The evolution of photoresponse in $H_2PP$ decorated $ReS_2$. a) The logarithmic plot of the normalized photocurrent under 50 pW, 0.5 nW, 50 nW and 5 μW. b) The photocurrent varies with the laser power contributed by the recombination of excess carriers and empty of shallow traps, respectively. c) The calculated percentage of photocurrent contributed by the slow process. d) Detectivity (D*) as a function of incident laser power.

# Supporting Information

**Defect engineering for modulating the trap states in two-dimensional photoconductor**

*Jie Jiang, Chongyi Ling, Tao Xu, Wenhui Wang, Xianghong Niu, Amina Zafar, Zhenzhong Yan, Xiaomu Wang, Yumeng You, Litao Sun, Junpeng Lu, Jinlan Wang[*] and Zhenhua Ni[*]*

## 1. Sample characterizations

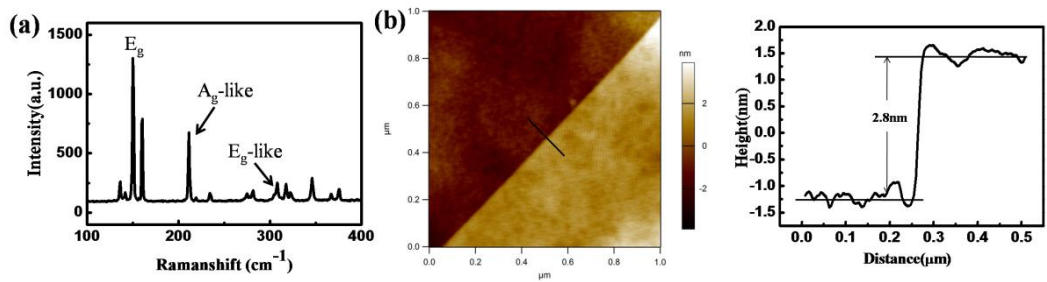

**Figure S1.** Sample characterization. a) Raman spectrum of pristine $ReS_2$. The three most prominent Raman peaks are at 150, 160 and 211 cm-1 respectively, correspond to the in-plane (Eg) and mostly out-of-plane (Ag-like) vibration modes, which is consistent with other report.[21] b) AFM image of pristine $ReS_2$. The thickness of ReS2 film is 2.8 nm, which is characterized by atomic force microscope (AFM), and corresponding to 3 layers (considering the thickness of monolayer $ReS_2$ of ~0.7 nm).[21]

## 2. Three procedures of photoconductivity decay in as-prepared $ReS_2$

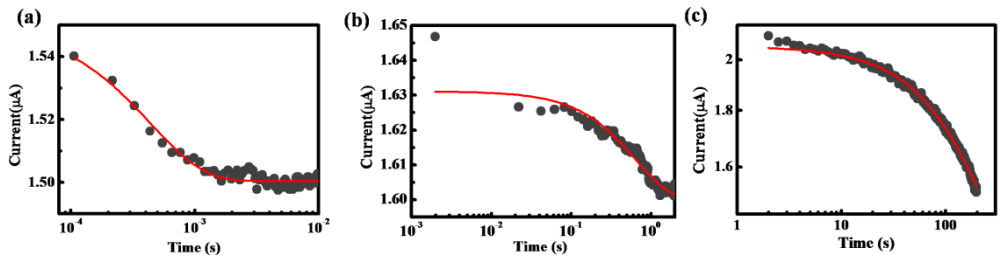

**Figure S2.** Three procedures of photoconductivity decay in as-prepared ReS$_2$. The photoconductivity decay curves of as-prepared ReS$_2$ in chopping frequency of a) 50 Hz b) c) without chopping.

There are mainly three procedures of photoconductivity decay in as-prepared ReS$_2$ device. When the chopping frequency is 50 Hz, the light-off time is only 10 ms, not enough for holes to escape from the traps. Therefore, the decay curve in Figure S2a only contains the recombination of excess carriers and the fitted decay time constant is ~0.4 ms. Figure S2b contains the empty of shallow traps (~0.6 s) in addition to the recombination process because negligible holes could escape from deep traps in the time scale of second. Figure S2c contains all three procedures: recombination, empty of shallow and deep traps (~200 s), respectively. We can see that the photoresponse of as-prepared ReS$_2$ device is dominant by deep traps.

### 3. Transmission Electron Microscope (TEM) measurement

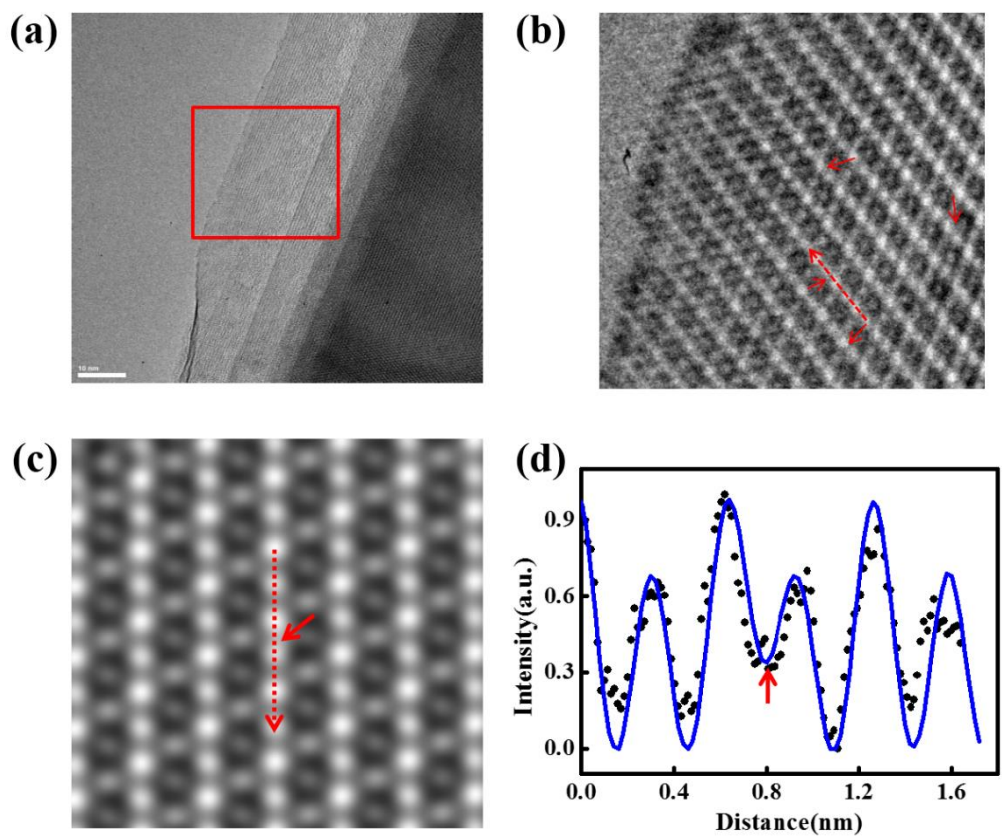

**Figure S3.** Transmission Electron Microscope (TEM) measurement. a) TEM of a 2 layer ReS$_2$ sample with S vacancies. b) High-resolution TEM image. c) Simulated high-resolution TEM image. d) Intensity profiles along the red dashed line in Figure S3b (black dot) and S3c (blue line).

To reveal the crystal structure and defects, we transfer the exfoliated ReS$_2$ samples onto the TEM grids and perform aberration-corrected TEM characterizations under 80 keV acceleration voltage. A TEM image of a 2-layer ReS$_2$ sample is shown in Figure S3a. Figure S3b shows the high-resolution TEM image of the sample in Figure S3a and Figure S3c is the simulated high-resolution TEM image. The S vacancies can be clearly distinguished by analyzing the intensity profile. Comparison between intensity profiles along the red dashed line in Figure S3b and S3c show quantitative agreement (Figure S3b), which confirms the existence of S vacancies.[19] The red arrows point out the S vacancies in our sample. S vacancies will introduce both shallow and deep traps. Therefore, as a 2D material containing both shallow and deep traps, ReS$_2$ is a suitable candidate to demonstrate the strategy of defect engineering for modulating the trap states, and to achieve a photodetector with both high responsivity and fast response.

**4. The distinctions between traps and recombination centers**

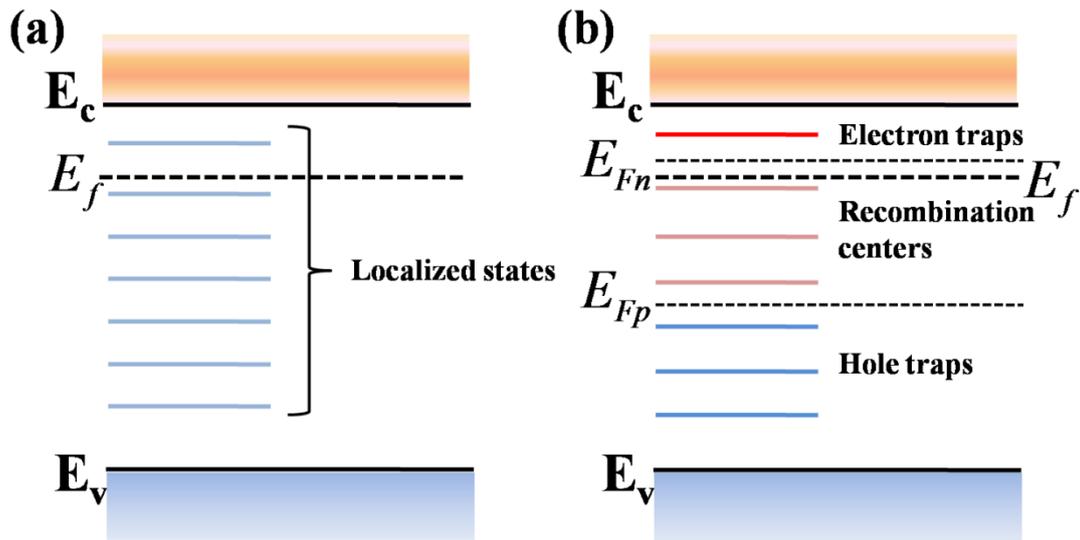

**Figure S4.** Schematic diagram of traps and recombination centers. a) In dark b) Under illumination.

The localized states may act as recombination or trap states depend on their position in bandgap.[6] Any states between the quasi-Fermi levels for electrons and holes is a recombination center which has been discussed by J. G. Simmons and G. W. Taylor.[27] A trap is amphoteric in the sense that it acts both as an electron trap and as a hole trap, depending on its state of occupancy. When the trap is empty, it is ready to receive an electron, and thus it is operating as an electron trap. When the trap contains an electron, it is ready to receive a hole, and hence is a hole trap. Under illumination, schematic diagram of recombination centers and traps is shown in Figure S4. The localized states above $E_{Fn}$ and under $E_{Fn}$ will then act as electron traps and hole traps, respectively.

## 5. The photoresponse of H$_2$PP decorated ReS$_2$ device

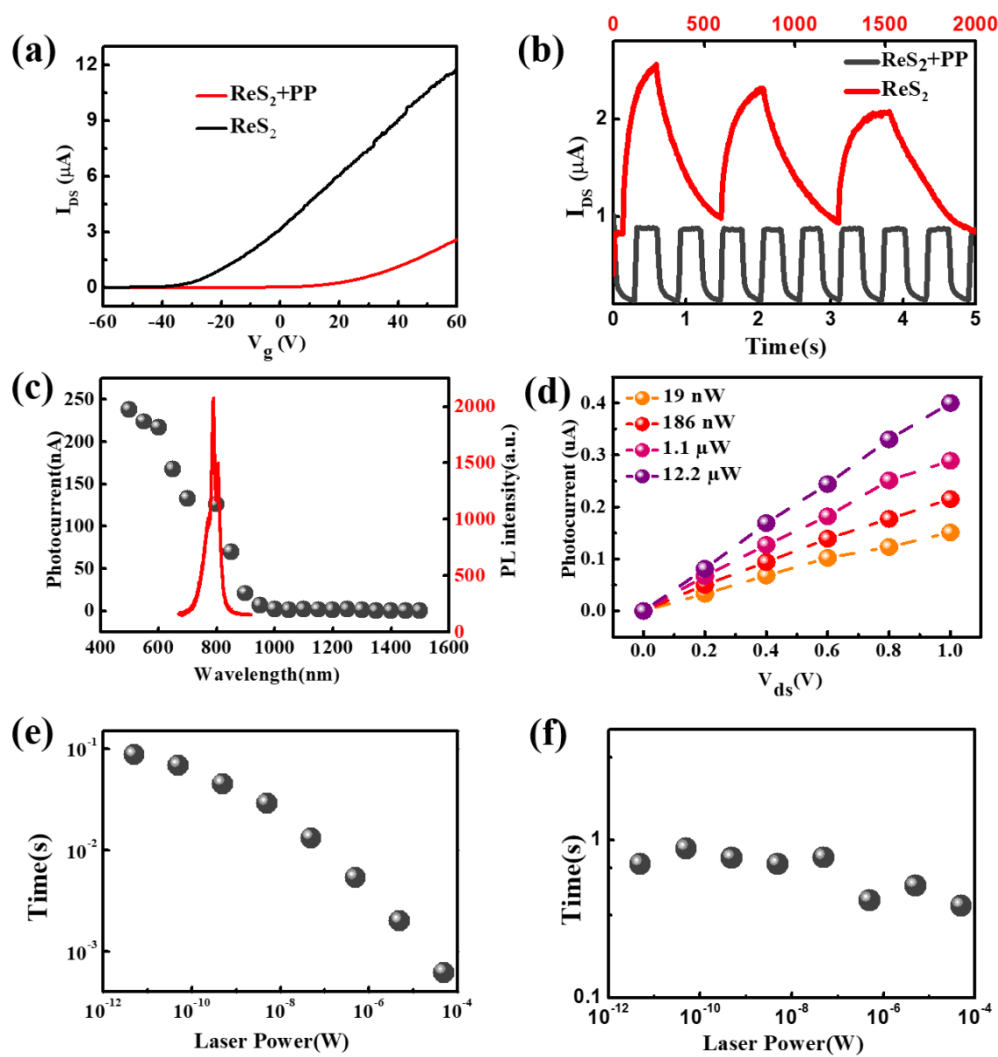

**Figure S5.** The photoresponse of H$_2$PP decorated ReS$_2$ device. a) Typical transfer curve, b) Photo-switching characteristics for as-prepared and H$_2$PP decorated ReS$_2$. c) Photocurrent evolution with laser wavelength. Laser power is fixed at 400nW. d) Photocurrent evolution with V$_{ds}$ at different laser power. e) The fast and f) slow components of the fall time extracted at different incident laser power.

Figure S5a presents the typical transfer curve of ReS$_2$ device before and after coating H$_2$PP molecules. The source–drain voltage (V$_{ds}$) is fixed at 1.0 V. There is a significant shift of the threshold voltage Vt after H$_2$PP coating, from ~-28.2 V to ~25.4 V, indicating a heavy p-type doping and electron transfer from ReS$_2$ to molecules. The density of transferred carriers is about $3.8 \times 10^{12}$ cm$^{-2}$, which can be estimated by $n = C_{ox}\Delta V_t / e$, where $C_{ox} = \varepsilon_0 \varepsilon_r / d$, $\Delta V_t$ is the shift of threshold voltage (53.6 V), e is the electron charge, $\varepsilon_0$ is the vacuum permittivity, $\varepsilon_r$ for SiO$_2$ is 3.9, and the thickness d of the SiO$_2$ is 300 nm. The electron field effect mobility is ~25.5 and ~13.1 cm$^2$/Vs, respectively. The mobility of H$_2$PP decorated ReS$_2$ device should be much higher than ~13.1 cm$^2$/Vs, because the device hasn't reached linear region at the gate voltage of 60V.

Figure S5b shows the photo-switching characteristics of the ReS$_2$ device with and without H$_2$PP molecules using laser (532 nm, spot size ~1 μm) focused on the channel under 1 V bias. For as-prepared ReS$_2$ device, the photoresponse is very slow, in the order of minutes, which is consistent with previous reports.[8, 9] Obviously, the response time is improved by several orders after H$_2$PP decoration. We exclude the possibility of retrapping effect for the relative symmetry of rising and decay curves and absence of long tail in decay signal, which are expected in the presence of retrapping.[6]

The spectral photocurrent response of the device is shown in Figure S5c, which is obtained by using a super-continuum light source with a tunable band pass filter. There is a strong decrease of photocurrent at 800-900 nm, and the photoresponse wavelength can be extended to ~1550 nm. The photoluminescence spectrum of multilayer ReS$_2$ (taken at ~83 K) is shown in Figure S5c (red curve), and there are strong band edge emissions at ~ 800 nm,[21] which is consistent with the spectrum photoresponse. This indicates that the photocurrent mainly originates from the absorption ReS$_2$, but not the absorbed molecules. The photoresponse at near infrared up to 1550 nm might be caused by defect absorption.[39] Figure S5d presents the photocurrent versus $V_{ds}$ under different laser power. The linear dependence indicates that photoresponse can be modulated by bias voltage, which is an attractive characteristic for imaging applications. The response time of recombination process and empty of shallow traps are plotted in Figure S5e and S5f, respectively. The carrier lifetime decreased with the increased laser power which is consistent with other reports.[5] The lifetime of the free excess carriers is inversely proportional to the density of free carriers.[1,2] When increasing the incident light power, the density of free carriers is increased by larger optical generation. The time constant then decreased significantly. According to $\tau_t^{-1} = s_p N_v V_{th} \exp(\frac{-\Delta E}{kT})$, the meaning trapping time (time of empty shallow traps) is independent of laser power.[22-24] The fitted times ranged from 0.3-0.9 s are constant indicates that the traps in $H_2PP$ decorated $ReS_2$ is a discrete level rather than a distribution in band gap.

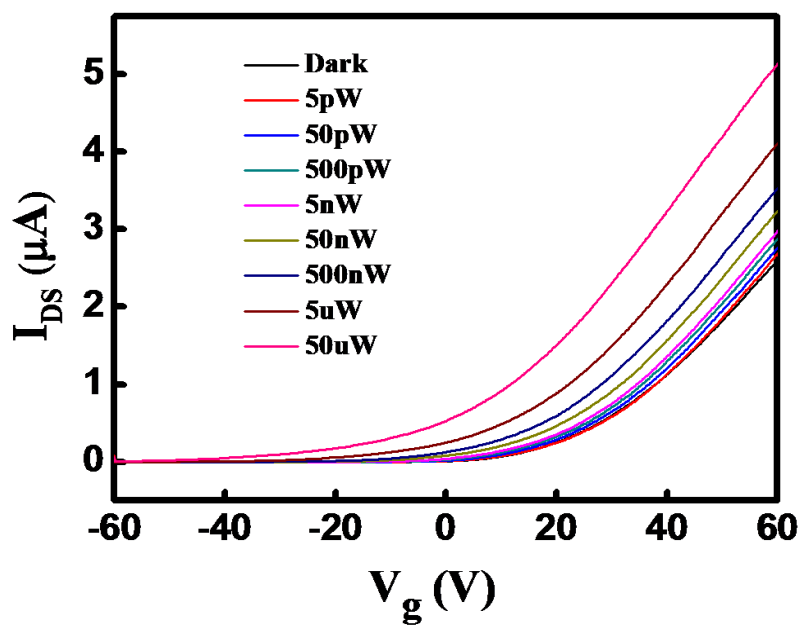

**Figure S6.** Transfer curve of $H_2PP$ decorated device under dark condition and different laser power.

The threshold voltage is moving to the negative direction of voltage, indicates the existence of hole trapping centers. These hole trapping centers could be S vacancies or $H_2PP$ molecules. The electrons transferred to $H_2PP$ molecules may also introduce photogating effect and increase the photoconductivity of our device. There is no effective method to distinguish defect trapping and photogating effect. However, the observed timescale of holes escaped from these trapping centers (0.1-1s) is in the same order of empty of shallow traps in as-exfoliated $ReS_2$ device, as shown in Figure S2b and S5f. Therefore, we conclude that the photoresponse of H2PP decorated device is dominant by defect trapping rather than photogating effect.

## 6. Stability of $H_2PP$ molecules on $ReS_2$

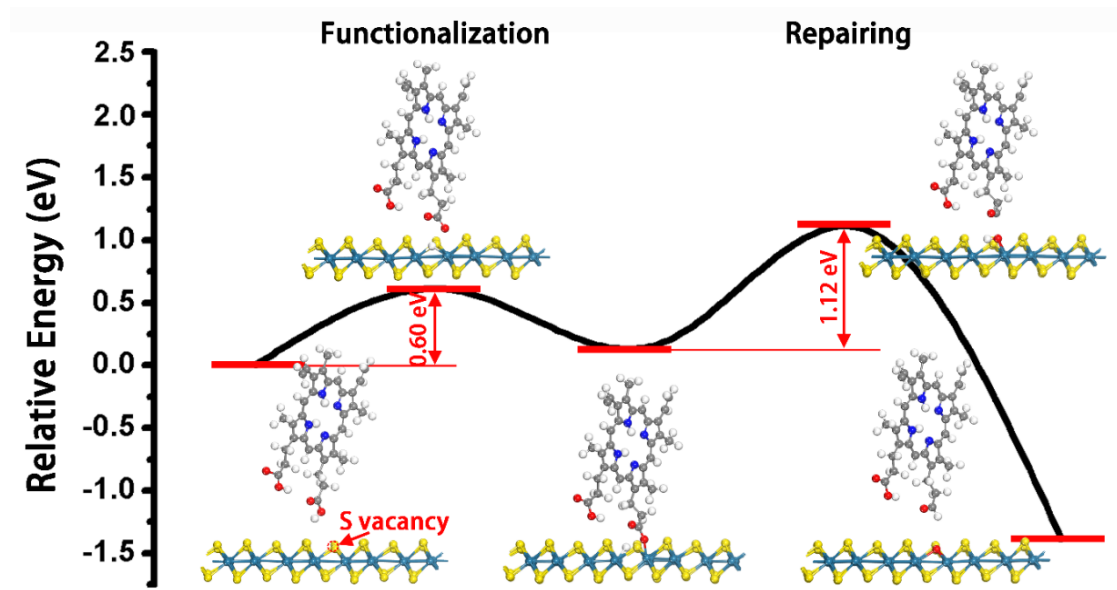

**Figure S7.** Kinetics and transient states of the reaction between a single S vacancy and a Protoporphyrin molecule through the functionalization and repairing mechanisms.

Generally, there are two possible reaction mechanisms between $H_2PP$ molecule and S vacancy: functionalization mechanism and repairing mechanism (Figure S7). As presented in the manuscript and Figure G, S vacancy in $ReS_2$ can be easily functionalized by $H_2PP$ molecule (functionalization mechanism) by overcoming a small energy barrier of 0.6 eV. The adsorbed $H_2PP$ molecule can further desorb from

ReS$_2$ surface via the breaking of C-O bond (repairing mechanism). However, this process needs energy injection of 1.12 eV (Figure S7). Therefore, the adsorbed H$_2$PP will be stabilized on ReS$_2$ rather than desorbs

## 7. Temperature dependent photoresponse measurement

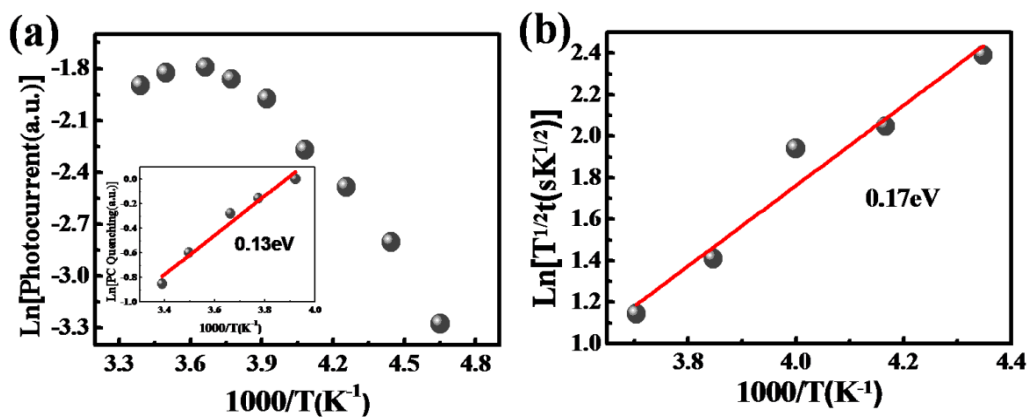

**Figure S8.** Temperature dependent photoresponse measurement. a) Temperature dependence of photocurrent and photocurrent quenching (shown in inset) reveals trap state with an activation energy of ~0.13eV. b) Photocurrent temporal response as a function of temperature reveals an activation energy of the sensitizing center of ~0.17 eV.

Temperature dependent of photoresponse in H$_2$PP decorated ReS$_2$ device is investigated. In photoconductors, the photocurrent is proportional to mobility and carrier lifetime. At low temperatures, all trapping centers are activated. Thus, carrier lifetime is independent of temperature.[22] Photocurrent then follows the mobility dependence on temperature. The relation of mobility with temperature can be described as:

$$\mu = \mu_0 \exp(-E_a / kT) \qquad (1)$$

Where $\mu_0$ is the temperature invariant mobility factor and $E_a$ is the mobility activation energy.[40] With increasing temperature, thermal energy is presumed to become sufficient to accelerate the emptying of the hole traps. Photocurrent in this regime is determined by the two competing mechanisms of increasing mobility and decreasing lifetime with temperature, leading to photocurrent quenching. The quenching rate is expected to be proportional to the following equation:

$$N_v / P_t \exp(-\Delta E / kT) \qquad (2)$$

Where $P_t$ is density of hole traps. Figure S8a shows the photoresponsivity as a function of temperature. At low temperatures, photoresponsivity increases with temperature following mobility thermal activation ($E_a$) of ~0.15 eV. Photocurrent quenching takes place as a result of thermal deactivation with increasing temperature. The slope of photocurrent quenching demonstrates an activation energy ($\Delta E$) of ~0.13 eV, as shown in inset of Figure S8a.

An additional way to determine the trapping level is to study the temperature dependence of the decay time.[22-24] According to $\tau_t^{-1} = s_p N_v V_{th} \exp(\frac{-\Delta E}{kT})$, the Arrhenius plot of $\tau T^{1/2}$ versus temperature also reveals the energy difference $\Delta E$ of the trap state and the band edge (Figure S8b). This method allows direct measurement of the trap state thermal emission rate and the trapping level is calculated to be ~0.17 eV above the valance band, which agrees reasonably well with the responsivity quenching results (~0.13 eV) and also theoretical calculation (~0.2 eV).

## 8. Noise current measurements

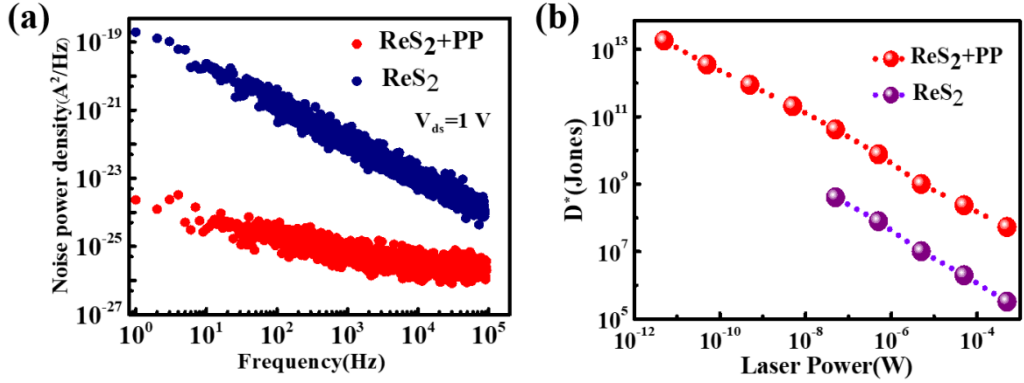

**Figure S9.** a) Spectra of the current noise power density of the as-prepared and $H_2PP$ decorated device. b) Detectivity (D*) as a function of incident laser power.

Noise power spectra $S_n(f)$ of a typical device before and after $H_2PP$ decoration were analyzed by using a noise characterization system (DPA inc. NC300) in ambient conditions. The sample was encapsulated in a metal box to be shielded from environmental noises. Figure S9(a) shows the spectra of the current noise power density of the as-prepared and $H_2PP$ decorated device plotted in double logarithmic coordinates, showing 1/f noise dominates. The specific detectivity $D^*$ can be calculated from the following equation:[32] $D^* = \frac{\sqrt{ABR}}{\sqrt{\langle i_n \rangle^2}} = \frac{\sqrt{ABR}}{\sqrt{\int_0^B S_n(f)df}}$ , where A is the area of device, B is the bandwidth, $\langle i_n \rangle^2 = \int_0^B S_n(f)df$ is the square noise current. For our device, A=23 μm² and B=100 kHz. $\sqrt{\langle i_n \rangle^2}$ were obtained to be $9.85 \times 10^{-9}$ and $5.15 \times 10^{-12}$ A for as-prepared and decorated device, respectively. The calculated D* of the device as a function of incident laser power ($P_{in}$) is then shown in Figure S9(b).

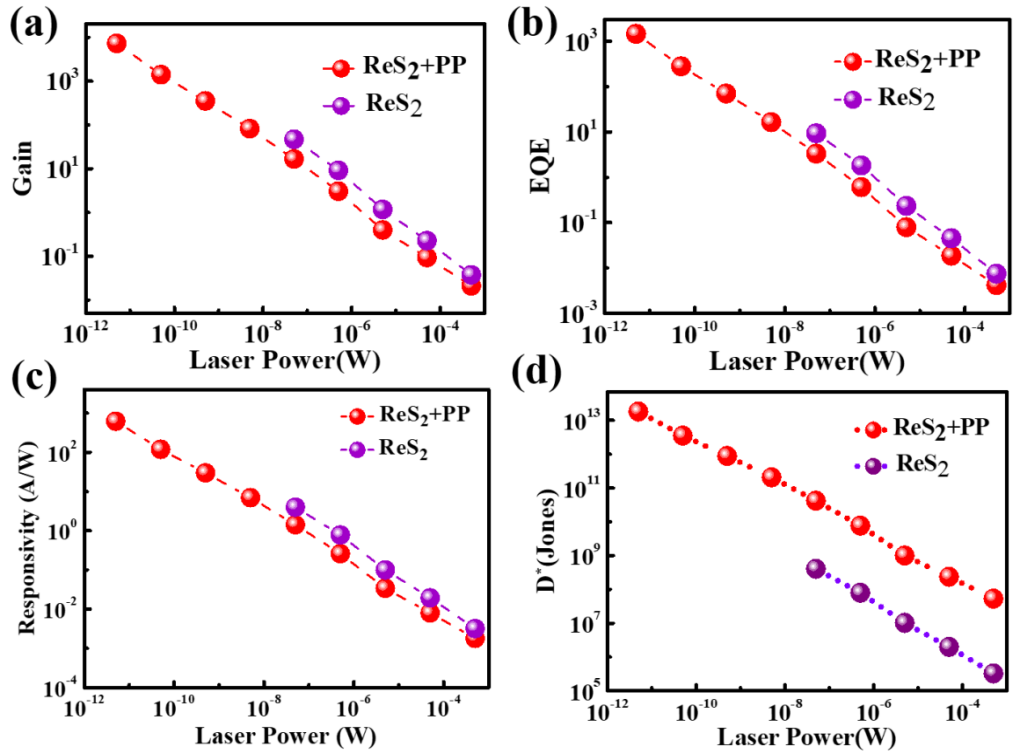

**Figure S10.** High sensitivity of $H_2PP$ decorates device. a) Photoconductivity gain b) External Quantum Efficiency (EQE) c) Photoresponsivity d) Detectivity of as–prepared and $H_2PP$ decorated device.

The EQE and responsivity of $H_2PP$ decorated device decreases slightly as compared to as-prepared device under the illumination of same laser power, as shown in Figure S10. However, due to the great reduction of current noise after $H_2PP$ decoration, the detectivity (D*) of the device is greatly enhanced. Therefore, the $H_2PP$ decorated device has much lower detection limit (pW) of light power.

## 9. Results from additional devices

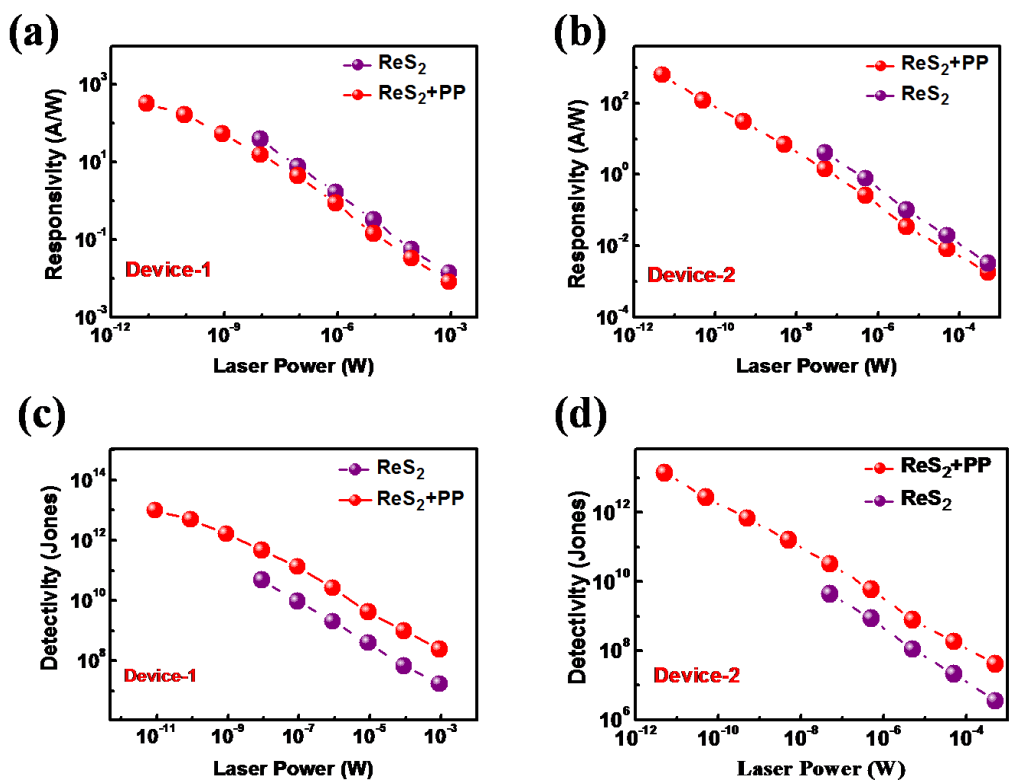

**Figure S11.** Photoresponse from additional devices. The photoreponsivity and detectivity of additional two devices before and after molecular decoration.

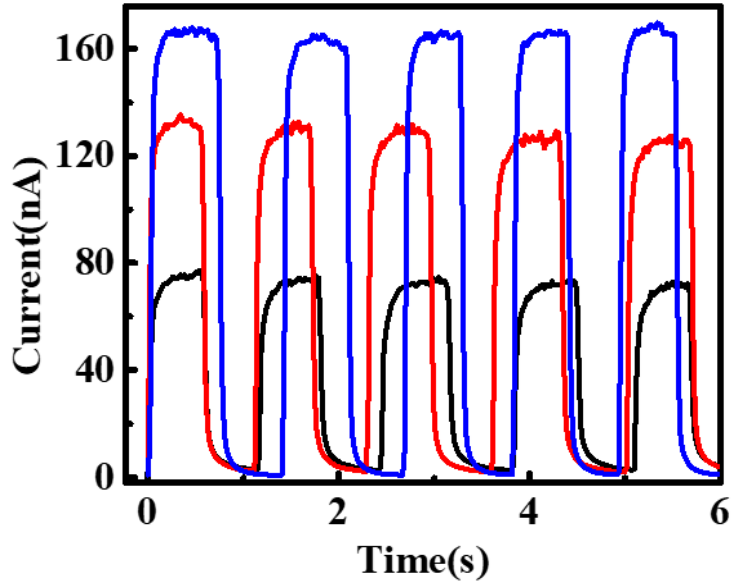

**Figure S12.** Typical Photo-switching characteristics for additional H$_2$PP decorated ReS$_2$ devices.

We have fabricated more than 10 devices with the layer number of ReS$_2$ ranges from 2-5 layers, and all the device show similar high responsivity and fast response. Figure 11 shows the photoreponsivity and detectivity of additional two devices before and after molecule decoration, where great improvement of detectivity can be observed. The detectivity calculated by the equation (4) is comparable with the value assuming that noise is dominant by dark current. Thus, here we use equation $D^* = \dfrac{A^{1/2}}{(2eI_d)^{1/2}} R$ to calculate the detectivity. Figure S12 shows typical photo-switching characteristics for additional H$_2$PP decorated ReS$_2$ devices.

## 10. The effects from remaining traps:

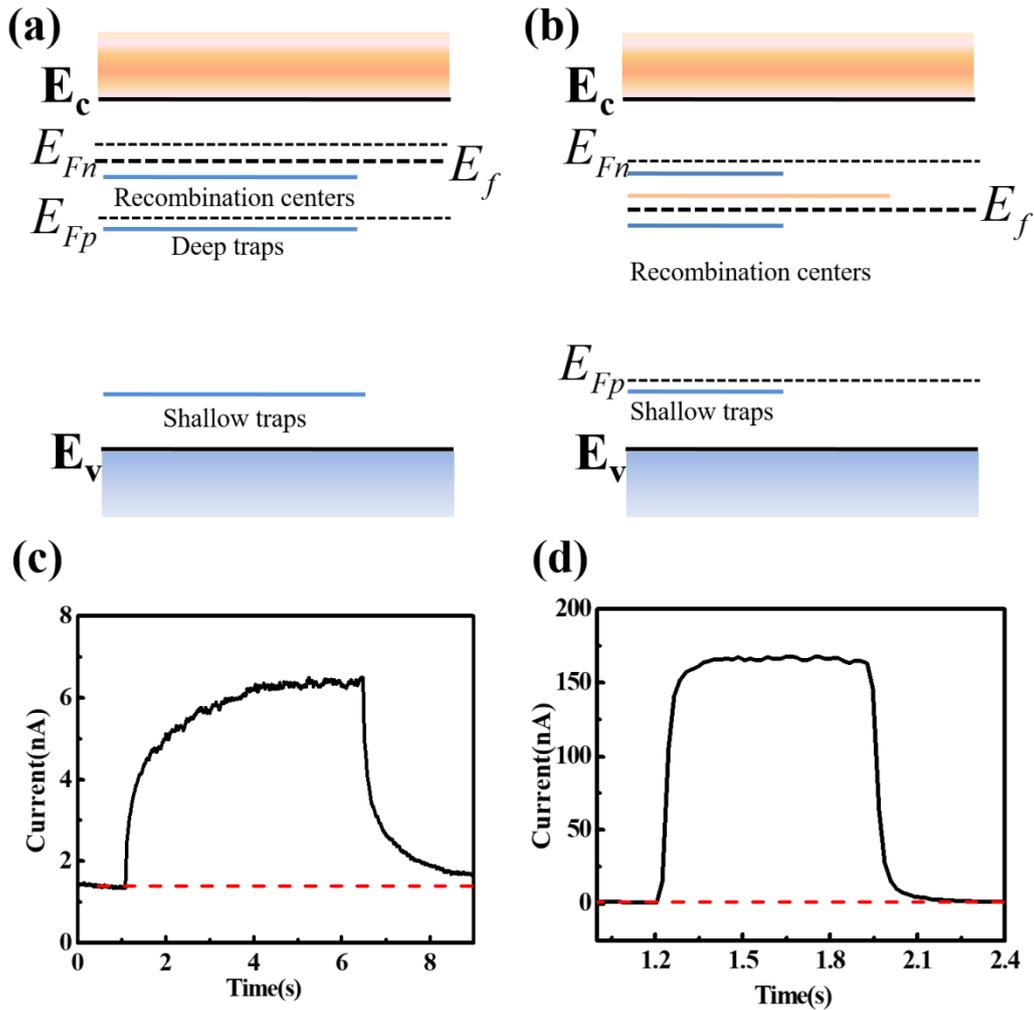

**Figure S13.** Effects from remaining deep traps. Schematic diagram of quasi Fermi level under illumination of a) as-prepared b) H$_2$PP decorated ReS$_2$. Transient response of H$_2$PP decorated ReS$_2$ at c) 5 pW d) 5 µW. The deep traps act as trap centers only in ultralow light intensity and convert into recombination centers in high light intensity.

In as-prepared ReS$_2$ device containing larger amount of defects, the quasi Fermi level for holes will be pinned near the deep trap states even under high illumination intensity,[31] as shown in Figure S13a. After H$_2$PP decoration, the healing of most S vacancies allows the quasi-Fermi level for holes to move down and shallow traps are activated (Figure S13b). The deep traps act as trap centers only in ultralow light intensity, and convert into recombination centers in high light intensity. At the time scale of seconds after stopping illumination, the device still has an increased photoconductivity caused by the filling of deep traps with a contribution of less than 10％ in very low power (5 pW), providing evidence that there is still a small amount of traps in H$_2$PP decorated ReS$_2$ (Figure S13c). The contribution from deep traps become negligible (<1%) with increased light intensity, as shown in Figure S13d. Obviously, the deep traps convert into recombination centers at high light intensity.